\documentclass[prl,aps,reprint,10pt,amsfonts,amssymb,amsmath]{revtex4-1}

\usepackage{xcolor}
\usepackage{graphicx}
\usepackage{bm}
\usepackage{dsfont}

\usepackage{latexsym}
\usepackage{amsmath,amsfonts,amssymb,epsfig,color}
\usepackage{amsthm}

\usepackage{subfigure}
\usepackage{color}
\usepackage[normalem]{ulem}

\def\R{\mathbb R}

\def\e{\varepsilon}

\def\R{\mathbb R}

\def\e{\varepsilon}

\newcommand{\beq}{\begin{equation}}
\newcommand{\eeq}{\end{equation}}

\newcommand{\medintinrigo}{-\kern -,315cm\int}
\newcommand{\medint}{-\kern -,375cm\int}

\begin{document}

\title{Engineering Curvature Induced Anisotropy in Thin Ferromagnetic Films}

\author{Oleg A. Tretiakov}
\email{olegt@imr.tohoku.ac.jp}
\affiliation{Institute for Materials Research, Tohoku University, Sendai 980-8577, Japan}
\affiliation{School of Natural Sciences, Far Eastern Federal University, Vladivostok 690950, Russia}
\author{Massimiliano Morini}
\affiliation{Universita di Parma, Parma, Italy}
\author{Sergiy Vasylkevych}
\affiliation{School of Mathematics, University of Bristol, Bristol, BS8 1TW, UK}
\author{Valeriy Slastikov}
\affiliation{School of Mathematics, University of Bristol, Bristol, BS8 1TW, UK}

\date{October 18, 2016}

\begin{abstract}
The large curvature effects on micromagnetic energy of a thin ferromagnetic film with nonlocal dipolar energy are considered. We predict that the dipolar interaction and surface curvature can produce perpendicular anisotropy which can be controlled by engineering a special type of periodic surface shape structure. Similar effects can be achieved by a significant surface roughness in the film. We show that in general the anisotropy can point in an arbitrary direction depending on the surface curvature. We provide simple examples of these periodic surface structures to demonstrate how to engineer particular anisotropies in the film. 
\end{abstract}

\maketitle

The puzzle of perpendicular magnetic anisotropy (PMA) origin in thin ferromagnetic films has a long history, dating back to N\'eel who was the first to address it \cite{Neel1954}.  Later, there have been several other attempts  in this direction  \cite{Bennett1971, Takeyama1976, Kolar1979, Bruno1988, Bruno1989}. In particular, in multilayers consisting of alternating ferromagnetic and heavy-metal (such as Pt) layers, PMA was attributed to strong spin-orbit interaction at the interfaces \cite{Daalderop1990, Bruno1989PRB, Kyuno1992, Wang1993PRL, Wang1993, Peng2015}.  However, in thin magnetic films PMA may exist without additional heavy-metal layer \cite{Wu_exp2015}, which enhances spin-orbit interaction in the system, thus pointing to a more general perpendicular anisotropy mechanism.  

Recent studies of magnetic structures with large-scale smoothly varying curvature have shown that the magnetization prefers to stay in tangential plane of the surface \cite{Gaididei2014, Carbou2001, Slastikov05_shells, Tretiakov2016_nanotube}. However, nowadays the thickness of magnetic films often reaches just a few monolayers, in which case the surface roughness may lead to large and rapid modulations of geometric curvature. This can significantly modify the magnetic properties and,  in fact, be responsible for the PMA in thin ferromagnetic films.

In this Letter we aim to understand the effect of surface roughness on the shape anisotropy and demonstrate the formation mechanism of perpendicular or any other given direction of magnetic anisotropy by means of surface engineering of a thin magnetic film. The proposed mechanism does not require any spin-orbit coupling and is related solely to the interplay of surface curvature and dipolar interactions in the film.  This possibility may open up a direction to tailor the interfacial magnetic anisotropy in thin ferromagnetic films without any additional layers of heavy metal, which, in turn, may lead to simpler and cheaper ways to engineer systems with any given anisotropy.  

Recently the curvature effects in thin magnetic films have become more accessible due to experimental advances in flexible electronics \cite{Streubel2014, Makarov2016, Bedoya-Pinto2014, LeeAPL2015, Makarov2016_review}, making the proposed method to control the anisotropy experimentally viable in the near future. Moreover, our findings imply that similar effects might be observed in thin films with significant surface roughness. 

\begin{figure}
    \includegraphics[width=\linewidth]{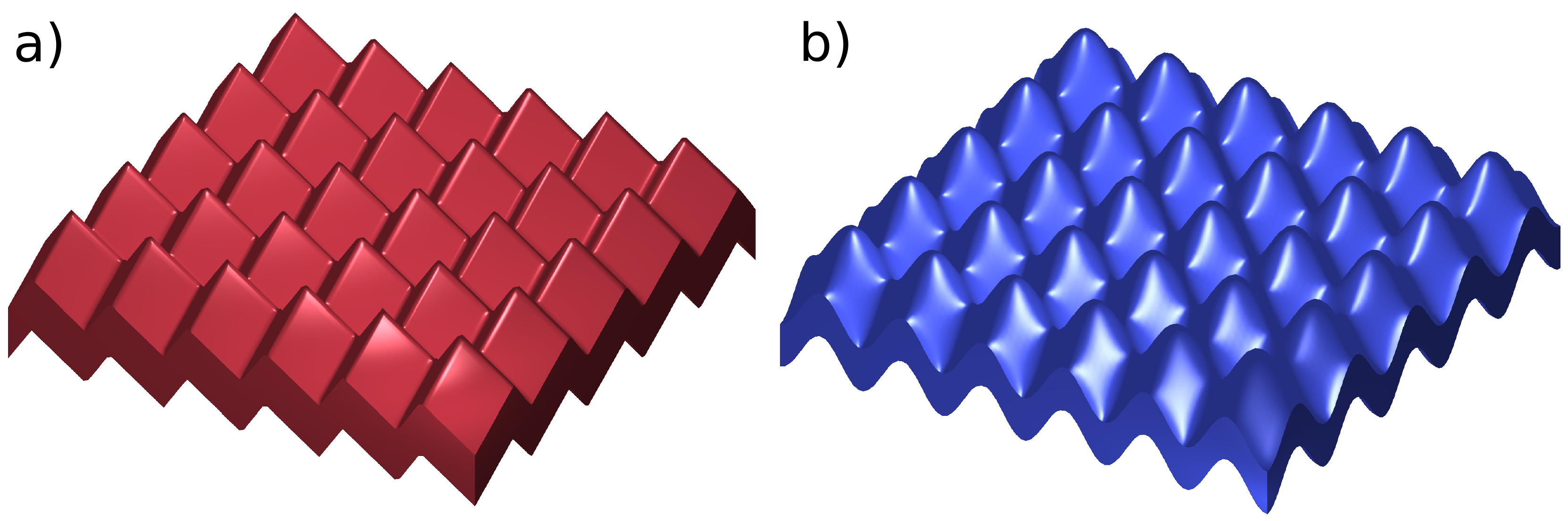}
 \caption{Examples of engineered periodic magnetic films. (a) The film consisting of pyramids. (b) The film consisting of $\sin^2 x_1 \sin^2 x_2$ shapes.}
 \label{fig:shapes}
 \end{figure} 

We obtain our results by employing asymptotic homogenization to tackle the physics of magnetic surfaces with small rapid periodic modulation. Physically the problem is associated with two scales, the larger scale is given by the size of the film's domain where we aim to determine the anisotropy, whereas the smaller one is given by the size of film's curvature modulation. The latter scale should be generally much smaller to have a nontrivial effect on anisotropy, which can be homogenized over the larger film's scale.  Formally, the method of asymptotic homogenization proceeds by introducing the fast variable $\mathbf{y}=\mathbf{x}/\e$  and performing an expansion of unit magnetization $\mathbf{M}$ in small parameter $\e$:
\begin{equation} 
\mathbf{M}_{\epsilon }({\mathbf{x}})=\mathbf{M}_{0}({\mathbf{x}})+\e \mathbf{M}_{1}({\mathbf{x}},{\mathbf{y}})+O(\e ^{2}),
\end{equation} 
which generates a hierarchy of problems. The homogenized equation is obtained and the effective coefficients are determined by solving the so-called "cell problems" for the function $\mathbf{M}_{1}({\mathbf{x}},{\mathbf{x}}/\e)$. For a thin film, parameter $\e=t/L$ is physically determined as the ratio of the film thickness $t$ and its typical lateral dimension $L$ (of the order of single domain size). 

We study a three-dimensional thin film domain
$V_\e =(\mathbf{x}', x_3)$, where $\mathbf{x}' = (x_1, x_2)$ belongs to the two-dimensional (2D) domain $\omega$ (the projection of the thin film on the plane) and $ \e f\left(\mathbf{x}'/\e \right)<x_3< \e [1 + f(\mathbf{x}'/\e )]$ with $\e >0$ being a constant dimensionless film's thickness. We consider an arbitrary periodic function $f(x_1,x_2)$, which models the film's surface modulation with the periodic cell given by a square of unit length.  Typical examples of the surface shape functions that might be considered are $f(\mathbf{x}')= \sin^2(\pi x_1)$,
$f(\mathbf{x}')= \sin^2(\pi x_1)\sin^2(\pi x_2)$, or the one shown in Fig.~\ref{fig:shapes} (a).

In the continuum description the dimensionless micromagnetic energy containing exchange and dipole-dipole interactions takes the form
\beq
\label{enMM}
{\mathcal E}_\e (\mathbf{M}) = \xi \int_{V_\e} |\nabla \mathbf{M} |^2\, d\mathbf{x} +  \int_{\R^3} |\nabla u |^2\, d\mathbf{x},
\eeq
where $\xi =A/(\mu_0 M_s^2 L^2)>0$ is the dimensionless material  parameter, $A$ is the exchange constant, $\mu_0$ is vacuum permeability, $M_s$ is saturation magnetization, and $u$ in the dipolar contribution is determined as the unique solution satisfying  
\beq\label{eqe}
\Delta u = {\rm div}\, \mathbf{M}   
\eeq
in the entire space, where the magnetization $\mathbf{M}$ is nonzero only within the volume of the film.

To study the thin film limit, i.e. the limiting behavior of the energy as $\e \to 0$, it is convenient to consider the rescaled energy $E_\e (\mathbf{m})={\mathcal E}_\e (\mathbf{M})/\e$ with magnetization  $\mathbf{M}(\mathbf{x}',x_3)=\mathbf{m}(\mathbf{x}',x_3/\e)$. 
In this limit, the main contribution to the energy is coming from the interaction of surface magnetic charges of the largest (top and bottom) surfaces. The limiting micromagnetic energy is given by the functional  
\begin{equation}
E_0 (\mathbf{m}) = \xi \int_{\omega} h^{\rm{ex}}(\nabla \mathbf{m})\, d\mathbf{x}'+\int_{\omega}K^{\mathrm{eff}} \mathbf{m}\cdot \mathbf{m}\, d\mathbf{x}',
\label{eq:E0}
\end{equation}
where a non-negative convex function $h^{\rm{ex}}$ vanishing at the origin is the exchange contribution   \footnote{See the Supplementary Material for the exact analytical expression.}, $K^{\mathrm{eff}}=[K^{\mathrm{hom}}+(K^{\mathrm{hom}})^{T}]/2 = \{ \kappa_{ij}\}$ is a symmetric  2nd-rank curvature induced effective anisotropy tensor with $i,j=1,2,3$, and 
the homogenized anisotropy matrix $K^{\mathrm{hom}}$ takes the form
\begin{equation}
\label{Ahom}
K^{\mathrm{hom}} = \frac{1}{2\pi}
\int d \mathbf{y}' \int_{\R^2} d \mathbf{z}' \, \mathbf{n}(\mathbf{y}') \otimes \mathbf{n}(\mathbf{z}'+\mathbf{y}') \left[  g(0) -   g(1) \right] 
\end{equation}
with
\begin{equation}
\label{g}
g(a) = \frac{1}{\sqrt{|\mathbf{z}'|^2 + \left|a+ f(\mathbf{z}'+\mathbf{y}') - f(\mathbf{y}') \right|^2}}  
\end{equation}
and 
$\mathbf{n}(\mathbf{y}')=(-\nabla f(\mathbf{y}'), 1)$. Here the integration over $\mathbf{y}'$ is performed in a unit square $[0,1]\times [0,1]$. We note that tensor $K^{\mathrm{eff}}$ is a non-negative definite, because the last term on the right-hand side of Eq.~(\ref{eq:E0}) is derived from the non-negative magnetostatic energy $\int_{\R^3} |\nabla u|^2$.

The main result of the Letter concerning the effective anisotropy behavior is based on Eqs.~(\ref{Ahom}) and (\ref{g}). In the following, we show that in thin magnetic films with periodic curvature one can engineer PMA or a uniaxial anisotropy of any particular orientation by choosing the surface shape $f(x_1, x_2)$ appropriately. This can open doors for tailoring the materials with a given anisotropy direction.  

\begin{figure}
    \includegraphics[width=0.85\linewidth]{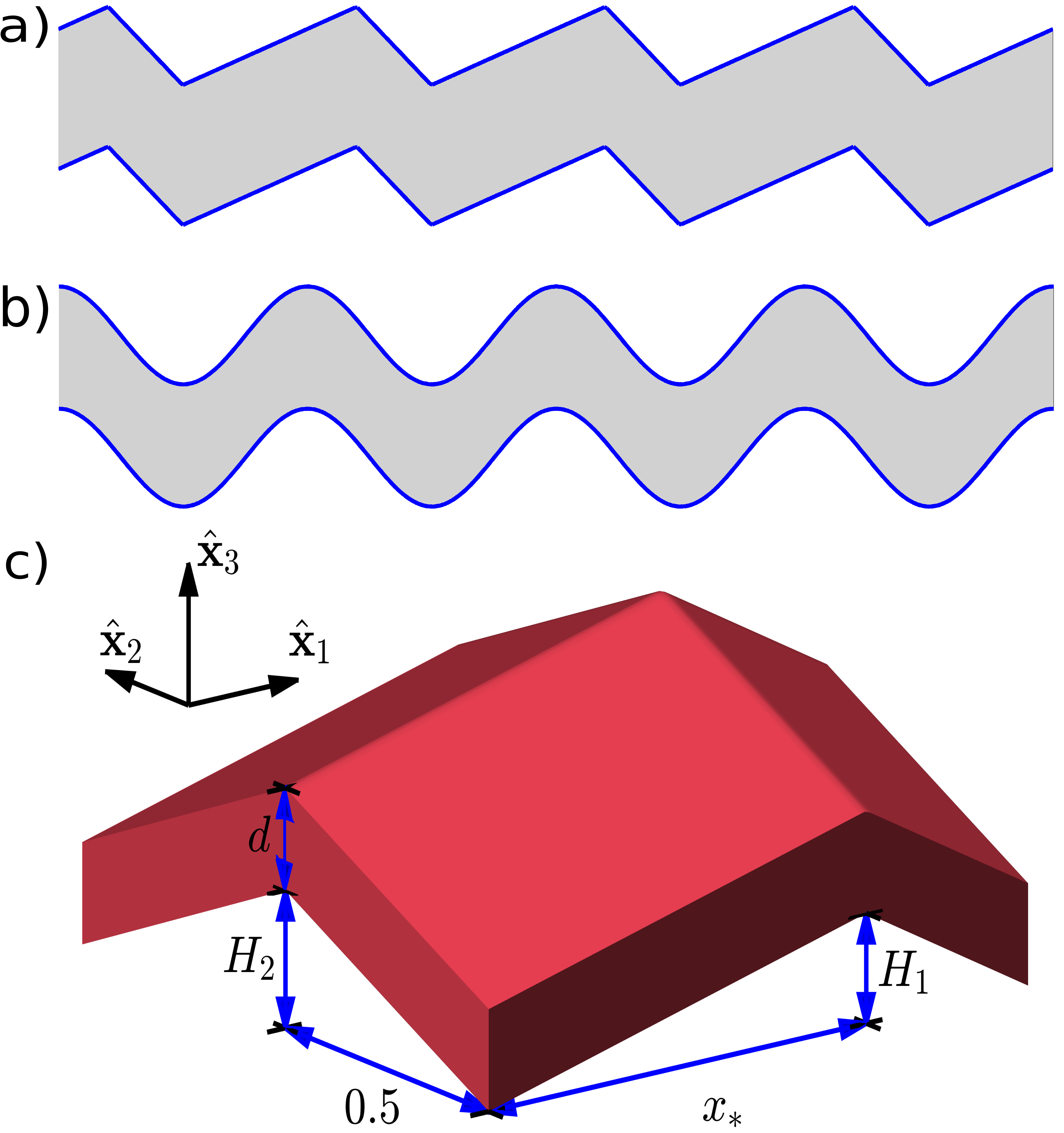}
 \caption{(a) -- (b) Cross-sections of thin films with periodic (a) pyramid and (b) $\sin^2 x_1 \sin^2 x_2$ textures. (c) One pyramid of the periodic structure shown in Fig.~\ref{fig:shapes}(a).}
 \label{fig:2Dshape}
 \end{figure}   
 
We first consider simpler, effectively one-dimensional (1D) case,  when the 2D structure changes periodically only in one direction (we choose this direction to be $\hat{\mathbf{x}}_1$).  In this case $f(x_1, x_2)=f(x_1)$ and after integrating Eq.~\eqref{Ahom}  over  $y_2$ and $z_2$, we obtain
\begin{eqnarray}
\label{2matrix1}
K^{\mathrm{hom}}_{\mathrm{1D}} &= &\frac{1}{4\pi}  \int_0^1 \!\!  dy_1 \! \int_{-\infty}^{\infty} \!\!\! dz_1 \, \mathbf{n}(y_1) \otimes \mathbf{n}(z_1+y_1) 
\nonumber \\
&&\times \log  
\frac{z_1^2 +[1+ f(z_1+y_1)-f(y_1)]^2}{z_1^2+[f(z_1+y_1)-f(y_1)]^2}.
\end{eqnarray} 
It is easy to check in Eq.~\eqref{2matrix1} that all effective matrix elements $\kappa_{i2}=\kappa_{2i}=0$ and therefore zero is guaranteed to be the minimal eigenvalue of $K^{\mathrm{eff}}$ with the eigenvector $\hat{\mathbf{x}}_2$.  If the zero eigenvalue is not degenerate, $\hat{\mathbf{x}}_2$ is the easy axis of anisotropy. Alternatively, if zero is an eigenvalue of multiplicity two,  the anisotropy is of easy $x_1x_2$-plane type. For example, this is the case when $f(x_1)=\mathrm{const}$. To conclude, for the structures periodically changing along $x_1$ direction, one can obtain only $x_1x_2$ easy-plane or easy-axis anisotropy along $\hat{\mathbf{x}}_2$.  

To explain the main ideas of how to engineer specific anisotropies, we next consider a truly 1D case by disregarding $\hat{\mathbf{x}}_2$ direction and investigating the anisotropy in the $x_1 x_3$-plane only. In this case $K^{\mathrm{eff}}_{\mathrm{1D}}$ is reduced to $2\times 2$ matrix 
\begin{equation}
\label{2matrix}
K^{\mathrm{eff}}_{\mathrm{1D}} = \begin{pmatrix}
\kappa_{11} & \kappa_{13} \\ 
\kappa_{13} & \kappa_{33} 
\end{pmatrix},
\end{equation}
and the anisotropy orientation is determined by its eigenvalues and eigenvectors. 
If the minimal eigenvalue of $K^{\mathrm{eff}}_{\mathrm{1D}}$,
\begin{equation}
\lambda_{\rm{min}}= (\kappa_{11}+\kappa_{33})/2-\sqrt{(\kappa_{11}-\kappa_{33})^{2}/4 +\kappa_{13}^2}, 
\end{equation}
is not degenerate (i.e. $\kappa_{11} \neq  \kappa_{33}$ or $\kappa_{13} \neq 0$), its eigenvector direction defines the easy axis of the anisotropy. Alternatively, if it is degenerate, the anisotropy is of $x_1x_3$ easy-plane type. For non-degenerate $\lambda_{\rm{min}}$, the anisotropy direction lies in $x_1x_3$-plane and makes angle $\phi$ with $\hat{\mathbf{x}}_1$:
\begin{equation}
\label{e.gamma}
\phi=
\operatorname{arctan} \left(\frac{1}{\gamma} \left[1-\operatorname{sgn}(\kappa_{33}-\kappa_{11})\sqrt{\gamma^2+1}\right] \right) 
\end{equation}
for $\kappa_{11}\neq \kappa_{33}$, where  $\gamma=2\kappa_{13}/(\kappa_{33}-\kappa_{11})$, 
 and in the special case $\kappa_{11}=\kappa_{33}$ and $\kappa_{13} \neq 0$ the angle $\phi= -\frac{\pi}{4} \operatorname{sgn}(\kappa_{13})$. 

\begin{figure}
    \includegraphics[width=0.99\linewidth]{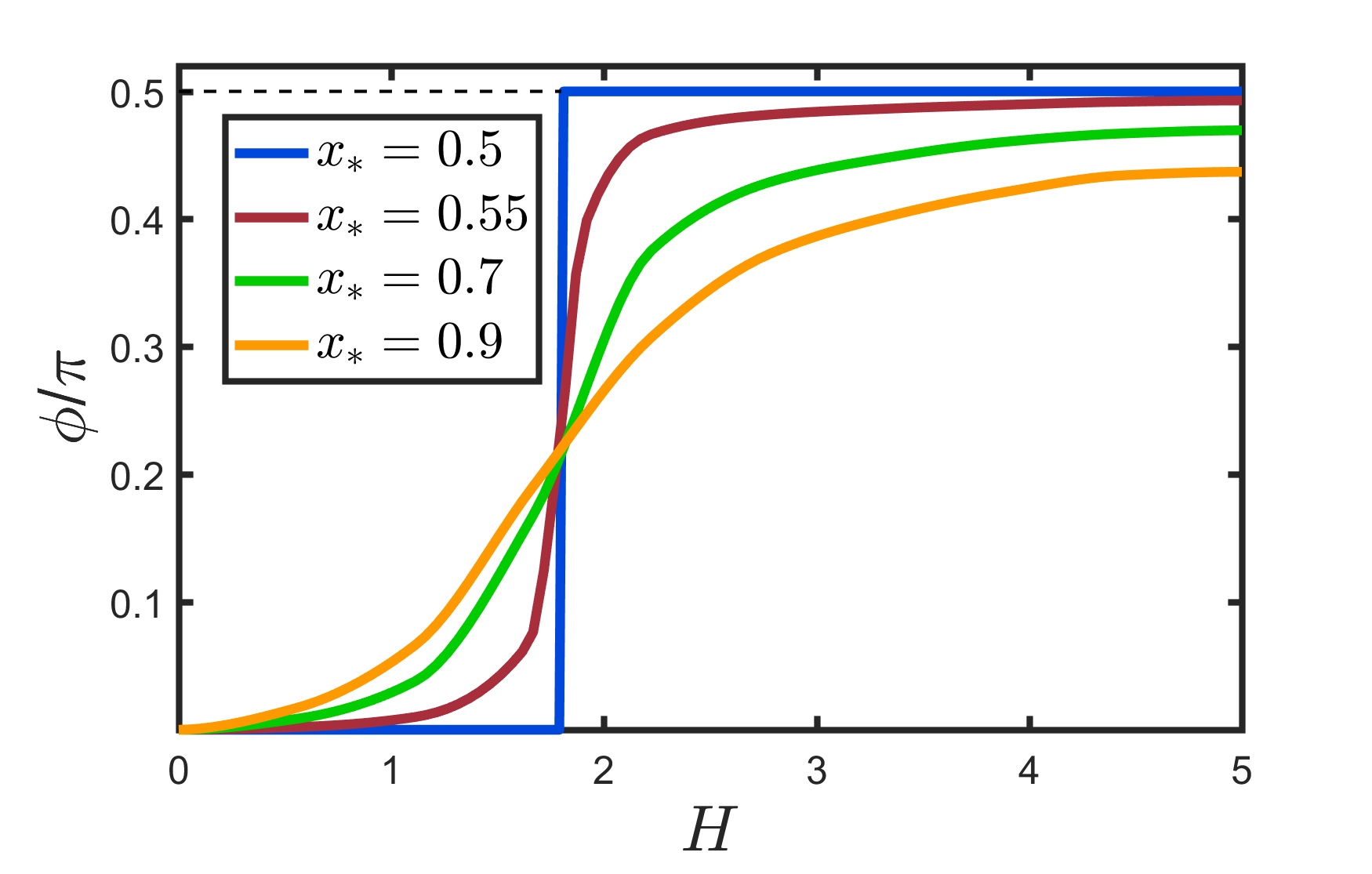}
 \caption{Angle $\phi$ the anisotropy makes with $\hat{\mathbf{x}}_1$ as a function of triangle's height $H$ in 1D case shown in Fig.~\ref{fig:2Dshape} (a).}
 \label{fig:diagram1D}
 \end{figure}

To be more specific we consider 1D films whose profile is given on each period by a triangle, see Fig.~\ref{fig:2Dshape} (a). Such a profile is completely characterized by the triangle's height $H$ and the position of the top vertex $x_{*}$, so on $[0,1]$ it is given by
\begin{equation} 
\label{e.eq1}
f_1(x_1; H,x_{*}) = \begin{cases} H\frac{x_1}{x_{*}}, & 0<x_1\leq x_{*}, \\ 
H\frac{1-x_1}{1-x_{*}}, &x_{*} <x_1 \leq 1. \end{cases} 
\end{equation}
One can show analytically that by varying $x_{*}$ and $H$ it is possible to align the easy-axis anisotropy with any direction in $x_1x_3$-plane. Here for simplicity we base our explanation on the results of numerical simulations presented in Fig.~\ref{fig:diagram1D}, where anisotropy angle $\phi$ as a function of triangle's height $H$ is calculated for the periodic 1D structure given by Eq.~(\ref{e.eq1}).  
First, we notice that setting $H=0$ results in the easy axis aligned with $\hat{\mathbf{x}}_1$. Figure~\ref{fig:diagram1D} shows that increasing triangle's height $H$, while holding $x_{*}>0.5$ fixed, continuously turns the easy axis from $0$ to $\pi/2$.   The special case of the symmetric triangles, $x_{*} = 0.5$, yields the easy-axis anisotropy along $\hat{\mathbf{x}}_1$ ($\phi =0$) below the critical value of triangle's height $H_c \simeq 1.8$ and along $\hat{\mathbf{x}}_3$ ($\phi =\pi/2$) above $H_c$. For $H=H_c$ the anisotropy is of easy-plane type. Thus, the range $[0,\pi/2]$ for the anisotropy angles can be covered by varying  $x_{*}$ in a reasonable range above $0.5$ and $H$ from zero to large enough $H>H_c$.  Now by changing parameter $x_{*}$ from $x_{*}>0.5$ presented in Fig.~\ref{fig:diagram1D} to $x_{*} <0.5$, due to the property $\phi(H,1-x_{*})=-\phi(H,x_{*})$, we can rotate the easy axis anisotropy by $\pi/2$ and cover the range $[-\pi/2, 0]$. As a result we conclude that in 1D case one can cover the entire range $[-\pi/2,\pi/2]$ of the easy-axis anisotropy orientations in $x_1 x_3$-plane.

Next we study a more general case of 2D structures modulated in both $\hat{\mathbf{x}}_1$ and $\hat{\mathbf{x}}_2$ directions. 
Two examples of these periodic structures, made of pyramids and $\sin^2 (x_1)\sin^2 (x_2)$ functions, are shown in Fig.~\ref{fig:shapes}. 
To be more specific and show the essential physics, we concentrate on a periodic structure made of pyramids depicted in Fig.~\ref{fig:2Dshape} (c).
Such a pyramid on a base of a unit square $[0,1]\times [0,1]$ with the apex located at $(x_{*}, 0.5)$  is modeled by the function
\begin{equation}
\label{func_f}
f(x_1,x_2; H_1,x_{*})=f_1(x_1; H_1, x_{*})+f_2(x_2; H_2, 0.5),
\end{equation}
where $f_{1,2}$ are given by Eq.~(\ref{e.eq1}). We choose in Eq.~(\ref{func_f}) the pyramid to be symmetric along $x_2$, because it will be sufficient to show the essential features by varying the asymmetry only along $x_1$. Since $f_2(x_2;H_2,0.5)$ is a symmetric function, i.e. $f_2(x_2)=f_2(1-x_2)$, the anisotropy matrix takes the form 
\begin{equation}
\label{e.Ahomsym}
		K^{\mathrm{eff}}=\begin{pmatrix}
		\kappa_{11} & 0 & \kappa_{13} \\
		0 & \kappa_{22} & 0 \\
		\kappa_{13} & 0 & \kappa_{33} 
		\end{pmatrix},
\end{equation}		
which is easy to show by exploiting the symmetries of  Eq.~\eqref{Ahom}. For Eq.~(\ref{e.Ahomsym}), $\hat{\mathbf{x}}_2$ is always an eigenvector of $K^{\mathrm{eff}}$ with the eigenvalue $\kappa_{22}$. Hence, the other eigenvalues and eigenvectors of $K^{\mathrm{eff}}$ are determined by analyzing the reduced matrix given by Eq.~(\ref{2matrix}) and have been already described in 1D case. 

\begin{figure}
    \includegraphics[width=\linewidth]{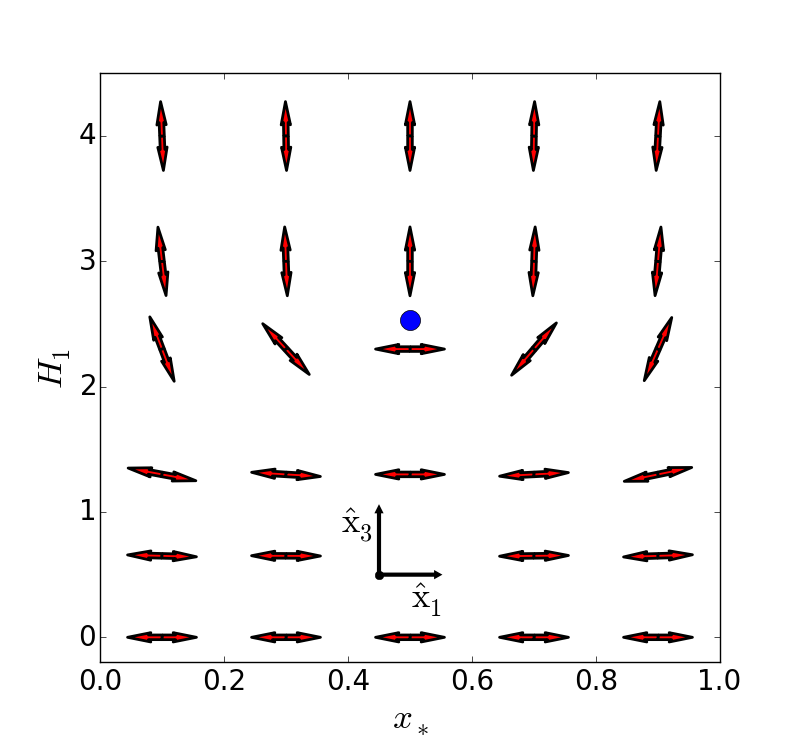}
 \caption{Direction of anisotropy in the film depicted in Fig.~\ref{fig:2Dshape}~(c) as a function of pyramid's height $H_1$ and position of the apex $x_{*}$ along
$\hat{\mathbf{x}}_1$. The film lies in the $x_1x_2$-plane and the base of each pyramid is a unit square.}
 \label{fig:anisotropy_D1_x0}
 \end{figure}

To confine the easy-axis to $x_1x_3$-plane, it is sufficient to choose the parameters so that either of the conditions $\kappa_{11} < \kappa_{22}$ or $\kappa_{33} < \kappa_{22}$ is satisfied.
For this we choose $H_2$ to be fixed and large enough, while $H_1$ and $x_{*}$ are allowed to vary.  Then, the anisotropy is of easy-axis type provided $\kappa_{11} \neq  \kappa_{33}$ or $\kappa_{13} \neq 0$. Using similar arguments as in 1D case, we can show that it is possible to cover the entire range of directions in the $x_1x_3$-plane. The corresponding results of numerical simulations using Monte-Carlo technique are presented in Fig.~\ref{fig:anisotropy_D1_x0} which shows the picture qualitatively identical to the 1D problem. We numerically observe that the value $H_2=5$ is large enough in the above discussed sense and use it in the simulations. The value of $H_c$, where for $x_{*}=0.5$ the anisotropy orientation abruptly changes from $\hat{\mathbf{x}}_1$ to $\hat{\mathbf{x}}_3$, is found to be $\simeq 2.52$ for $H_2=5$; it is shown by the blue point in Fig.~\ref{fig:anisotropy_D1_x0}. 

In order to obtain the preferred anisotropy in any direction it is enough to rotate pyramids by an appropriate angle in the $x_1x_2$-plane and repeat the same arguments as above. Analogous results can be obtained for smooth 2D functions such as $\sin^2 (\pi x_1)\sin^2 (\pi x_2)$ shown in Fig.~\ref{fig:shapes} (b), etc.

\textit{Discussion.}- 
The result of this Letter shows that, in spite of conventional belief \cite{Gay1986} that the dipole-dipole interaction in films thicker than a monolayer would put the magnetization in the plane of the film, in the case of particular surface modulation (or periodic roughness) this interaction can lead to perpendicular or any other uniaxial anisotropy. We note that the problem considered above with the same periodic profile on both surfaces can be extended even further. In the Supplementary material we provide the result derived for the more general case, where bottom and top profiles of the film are different \footnote{See the Supplementary Material for the expression and more details.}. Moreover, analogous homogenization technique may be used to treat two coupled magnetic films with periodically modulated surfaces  \cite{Neel_orange_peel_62, Parkin2000}, this problem will be treated elsewhere \cite{Slastikov_future}.

A possible experimental confirmation of our findings is corroborated by recent observation of the giant enhancement of magnetic anisotropy in ultrathin (6 nm La$_{0.67}$Sr$_{0.33}$MnO$_{3}$ films grown on (001) SrTiO$_{3}$ substrates) manganite films via nanoscale 1D periodic depth modulation, where the top 2 nm were patterned into periodic stripes \cite{Rajapitamahuni2016}. Generally, the magnetic systems studied in Ref.~\cite{Ball2014} may be excellent candidates for the curvature-induced anisotropy engineering.

The applicability limits of the asymptotic homogenization theory presented here are set by two scales: 1) the lower bound is determined by the validity of continuos model, i.e., it works on scales larger than interatomic spacing,  2) the upper bound is given by the scale of a single domain. Additionally,  since the main complexity in determining the magnetic anisotropy is associated with understanding the influence of the magnetostatic energy, which is non-local, without loss of generality our method can be extended to additively include local terms such as Zeeman energy and crystalline anisotropy.

In summary, we have demonstrated that the perpendicular anisotropy can be achieved in thin ferromagnetic films solely due to an interplay of surface curvature and dipolar interactions. This points to the fact that the surface roughness may significantly modify anisotropy.  We have shown how the nonlocal in their nature dipolar interactions, in the presence of arbitrary large surface curvature of the film, can be reduced to local effective anisotropy term in the magnetic energy. We modeled the film's surface shape by simple smooth functions $f(x_1,x_2)$, which can, in principle, be engineered in the films and demonstrated that by an appropriate choice of $f{(x_1,x_2)}$, one can orient the magnetic anisotropy axis along any direction. This provides a justification of a concept for future magnetic film nano-engineering with any chosen anisotropy without additional need of heavy-metal layers to provide spin-orbit coupling effects. This method would also allow to simplify the magnetic structures, by limiting them to only one magnetic layer.  

O.A.T. acknowledges support by the Grants-in-Aid for Scientific Research (Grants No. 25800184, No. 25247056, and No. 15H01009) from the Ministry of Education, Culture, Sports, Science and Technology (MEXT) of Japan and SpinNet. V.S. and S.V. acknowledge support by the EPSRC Grant EP/K02390X/1.

\bibliography{./anisotropy_curvature}

\end{document}